\documentclass[12pt]{article}
\usepackage{amssymb,amsmath}

\newcommand{\be}{\begin{equation}}
\newcommand{\ee}{\end{equation}}
\newcommand{\ba}{\begin{eqnarray}}
\newcommand{\ea}{\end{eqnarray}}

\def\laq{\raise 0.4ex\hbox{$<$}\kern -0.8em\lower 0.62ex\hbox{$\sim$}}
\def\gaq{\raise 0.4ex\hbox{$>$}\kern -0.7em\lower 0.62ex\hbox{$\sim$}}
\begin{document}

\centerline{\Large\bf The return of the membrane paradigm?}

\centerline{\Large\bf Black holes and strings in the water tap}
\bigskip
\bigskip
\centerline{{\bf Vitor Cardoso} \footnote{email: vcardoso@phy.olemiss.edu}\footnote{Also at Centro de
F\'{\i}sica Computacional, Universidade de Coimbra, P-3004-516 Coimbra, Portugal}} \centerline{Department of
Physics and Astronomy, The University of Mississippi, University, MS 38677-1848, USA }
\bigskip
\centerline{{\bf \'Oscar J. C. Dias} \footnote{email: odias@ub.edu}}
\centerline{Departament de F\'{\i}sica Fonamental, Universitat de
Barcelona, Av. Diagonal 647, E-08028 Barcelona, Spain}
\bigskip
\centerline{{\bf Leonardo Gualtieri} \footnote{email: gualtieri@roma1.infn.it}} \centerline{Dipartimento di Fisica, Universit\`a di Roma ``Sapienza''
and Sezione INFN Roma1, P. A.Moro 5, 00185, Roma, Italy }
\bigskip
\bigskip
\bigskip

\begin{abstract}
Several general arguments indicate that the event horizon behaves as
a stretched membrane. We propose using this relation to understand
gravity and dynamics of black objects in higher dimensions. We
provide evidence that (i) the gravitational Gregory-Laflamme
instability has a classical counterpart in the Rayleigh-Plateau
instability of fluids. Each known feature of the gravitational
instability can be accounted for in the fluid model. These features
include threshold mode, dispersion relation, time evolution and
critical dimension of certain phase transitions. Thus, we argue that
black strings break in much the same way as water from a faucet
breaks up into small droplets. (ii) General rotating black holes can
also be understood with this analogy. In particular, instability and
bifurcation diagrams for black objects can easily be inferred. This
correspondence can and should be used as a guiding tool to
understand and explore physics of gravity in higher dimensions.
\end{abstract}

\newpage

\section*{Introduction}

Gravity in higher dimensions has a long and tortuous story, beginning with the first attempts by Kaluza and
Klein to unify gravity and electromagnetism. Recent proposals include large extra-dimensional models and
braneworlds, advocated as a possible solution to the hierarchy problem of gauge couplings. Perhaps the most
famous example of higher dimensional theories is string theory, in which the previous scenarios can be
embedded. Higher-dimensional theories generally possess more degrees of freedom, thus providing a richer arena
to describe physical phenomena. However, this extended freedom has a cost: the parameter space to be searched is much
wider and calculations in higher dimensional gravity are technically very challenging. Thus a case-by-case
search has to be performed, with the associated time-costly computations. Moreover, it is not uncommon that the
necessary mathematical machinery is not even available, making the problem altogether impossible to handle.

It is thus of the utmost importance to sidestep the unwanted complex technicalities and infer or identify
directly the general, physically important features of the problem. A very popular approach to bypass hard
technical computations is to use analogue models \cite{bohrwheeler,plateau,membraneparadigm}: one builds on
previously established, well understood results, usually concerning a completely different setup, to infer the
general behavior of the problem at hand. The power of drawing analogies is that they enable one to predict and
gain {\it intuition}.

What we propose here is to use {\it fluids held together by surface tension} as models for gravitational
objects in a general number of spacetime dimensions \cite{Cardoso:2006ks,Cardoso:2006sj}. In simple terms, this
means that some of the effects seen for instance in water or soap bubbles should have a natural counterpart in
the gravitational sector. The idea of using analogues of this kind to extract useful information is not new:
more than 200 years ago Plateau \cite{plateau} conducted many experiments on liquid drops. The purpose of such
studies was to model giant liquid masses held together by self-gravitation (planets and stars) using
centimeter-sized drops held together by surface tension. Another well-known case, where the fluid drop model
proved extremely useful, is Bohr and Wheeler's \cite{bohrwheeler} description of nuclear fission as the rupture
of a charged liquid drop, where now the surface tension plays the role of nuclear forces. A final widely known
example, closely related to this essay, is the ``membrane paradigm'' for black holes by Thorne et al
\cite{membraneparadigm}. This paradigm establishes a rigorous mapping between the properties of the event
horizon and those of a stretched membrane, endowing the former with well-defined mechanical and electromagnetic
properties.

There is an extra motivation to relate the  properties of event
horizons to fluids with surface tension. Take the first law of black
hole mechanics \cite{BardCartHawk} describing how a (for simplicity,
uncharged, static) black hole, characterized by its mass or energy
$E$ and horizon area $A$, evolves when we throw an infinitesimal
amount of matter into it:
\be dE=T dA.\label{firstlaw}\ee
This law is nothing but a manifestation of the thermodynamical properties of black holes: they can be
ascribed a surface gravity proportional to its temperature $T_{\rm Hawking}=4T=1/(8 \pi E)$. Alternatively, as
first argued by Smarr \cite{smarr}, Eq. (\ref{firstlaw}) can be interpreted as a law for fluids, with $T$ being
an effective surface tension \cite{booktension}. This is rather intuitive: in fluids the potential energy is
associated with the storage of energy at the surface, therefore proportional to the area. The statement
that, for a given energy, the black object prefers the configuration with more entropy translates to the
well-known hydrodynamic property that, for a given volume, the fluid picks the configuration with less surface
area.

Our purpose here is to elaborate on the relationship between event horizons and membranes in higher dimensional
arenas. We show that with this correspondence at hand, many of the properties of black objects in higher
dimensions can be derived with little effort. It is also an extremely powerful tool to predict new phenomena
and in general to allow one to {\it understand} gravity in higher dimensions.

\section*{Rayleigh-Plateau and Gregory-Laflamme instabilities}

An illuminating example of the power of this analogy concerns black strings and branes
\cite{horowitzstrominger}. These are extended black holes: the horizon, instead of having the
topology of a sphere, can have for instance the topology of sphere times a line -- a cylinder. The simplest
black string is described by the metric of a $D$-dimensional Schwarzschild black hole times a line. Black
strings are unstable against gravity, in a mechanism known as Gregory-Laflamme (GL) instability \cite{gl}. The
GL mechanism makes any small perturbation with wavelength $\lambda$ of the order of, or larger than, the radius
of the cylinder $R_0$ grow exponentially with time. For wavelengths larger than a threshold $\lambda_c$,
$k_c R_0\equiv 2\pi R_0/\lambda_c \sim \sqrt{D} \label{GLkol}$
(for large $D$), the instability appears \cite{blackstringsreview}.
In this essay, we propose the dual object to the black string to be a fluid cylinder held by surface tension.
As shown by Plateau \cite{plateau}, a cylinder longer than its circumference is energetically unstable to
breakup: any axisymmetric small disturbance decreases the surface area of the cylinder. This is known as the
Rayleigh-Plateau (RP) instability. Indeed, consider a disturbed hyper-cylinder, with $D-1$ spatial directions,
radius $R_0$ and a transverse direction $z$. A simple calculation \cite{Cardoso:2006ks} yields $A=A_0\left(
1+\frac{\epsilon^2R_1^2}{4R_0^2}\left[k^2R_0^2-(D-2) \right ]\right )$ for the area of the disturbed cylinder,
with $A_0$ the undisturbed area. The potential energy per unit length is therefore $ P\propto
\left[k^2R_0^2-(D-2)\right]T$. We conclude that the system is unstable for $k_c R_0<\sqrt{D-2}$, since in this
case the perturbation decreases the potential energy. Note the remarkable quantitative agreement between the
threshold modes of the RP and GL instabilities for large number of spatial dimensions $D$.

The agreement holds for all known features of these instabilities
\cite{Cardoso:2006ks}. The dispersion relation (mode frequency {\it
vs} wavenumber) of the two instabilities are quite similar and have
the same dependency on space dimension $D$ \cite{Cardoso:2006ks}.
Non-axisymmetric modes are stable both in the RP and GL
instabilities. The existence of a critical dimension for both
objects provides an extra non-trivial check on the duality we are
proposing: one finds that, for $D\leq D_{\rm c}$ (with $D_{\rm
c}=11$), a spherical configuration has less surface area than the
cylinder and is thus a favored endpoint \cite{Cardoso:2006ks}. For
$D>D_{\rm c}$ this is no longer true. A similar critical dimension
is present in the GL side \cite{blackstringsreview}: for $D<D_{\rm
c}$ ($D_{\rm c}=13$) the black string is entropically unstable
against the formation of a spherical black hole.  If rotation is
added to the fluid, the strength of the fluid instability increases
due to centrifugal force effects. Thus, the proposed duality
predicts that even extremal black strings should be unstable. This
was very recently confirmed in the gravitational sector
\cite{Kleihaus:2007dg}.

\subsubsection*{The fate of black strings and their duals}

The full time evolution of the RP instability is well known
(numerically and experimentally \cite{Eggers}) while so far, only
the initial stage of the GL has been numerically studied
\cite{blackstringsreview}. The available results are consistent with
the duality viewpoint: starting from a single sinusoidal
perturbation both develop an almost cylindrical thread or neck in
between the two half rounded boundary regions. These agreements
raise of course several questions. The first is related to the
endpoint of the instability. What can we say about it, what insights
can the analogue or effective model offer? The RP instability makes
a cylinder pinch off. We end up with an array of main and satellite
drops with different sizes with less surface area than the initial
cylinder \cite{plateau,Eggers}. The analogy then suggests that a
similar configuration should exist as a static solution on the
gravity side. Very recently, this configuration of multi-black holes
with different masses was constructed perturbatively as a solution
of Einstein gravity \cite{Dias:2007hg}. In analogy with the fluid,
we can understand this gravitational endpoint as due to the fact
that the array of black holes has more entropy than the initial
uniform black string.

\subsubsection*{A critical dimension for phase transitions of black strings and fluids}

The threshold mode in the GL analysis signals a bifurcation to a new {\it static} branch of non-uniform strings
\cite{gubser}. A study by Sorkin \cite{blackstringsreview} concluded that for $D<D_*$ (with $D_*\sim 12,13$)
the phase transition from uniform to non-uniform black strings is of first order, while for larger $D$ it is
second order, therefore smoother. Within the fluid dual there is a similar static branch and critical dimension
$D_*$ for the phase transition. An heuristic, back-of-the-envelope computation can be done as follows. Take a
fluid cylinder, and expand the geometry around the zero mode (presented above) of a static uniform cylinder.
Specifically, consider the parametrization $r(z)=R_0+\epsilon \,R_1\cos (k_c z)+\epsilon^2 R_2 +\epsilon^3
R_3\cos (3Kz)$, where we set $K=k_c+\epsilon k_1$. The factors $R_2$ and $k_c$ were determined in
\cite{Cardoso:2006ks} by requiring volume conservation and area extremization: $R_2=-\frac{D-2}{4R_0}R_1^2$ and
$R_0 k_c=\sqrt{D-2}$. Going to next order ($\epsilon^3$), and demanding that both the uniform and the
non-uniform cylinder have the same volume, we find $R_3=\frac{(D-1)(D-2)^{3/2}(D-4)}{64k_1R_0^4}R_1^4$, $k_1$
is left undetermined, and
\be \frac{A_{\rm non-uniform}-A_{\rm uniform}}{A_{\rm
uniform}}=\frac{(D-2)(D-4+\sqrt{10})(D-4-\sqrt{10})}{64R_0^4}\epsilon^4 R_1^4\,. \ee
We conclude that at $D=D_*\sim 7.2$ there is a change in the smoothness of the phase transition; in this case
the transition seems to be smoother for $D<D_*$. It would be very interesting to perform a more rigorous
analysis of this transition, within a mathematically rigorous framework.

\section*{Rotation and bifurcation: black holes and black rings}

The correspondence proposed here can improve -- and strengthen -- our understanding of other gravitational
objects in higher dimensions. Take for instance spherical black holes. Their fluid analogues -- liquid drops --
are solutions of fluid dynamics in higher dimensions. The evolution of these objects once rotation is added was
reported in \cite{Cardoso:2006sj}. The spherical drop (for zero rotation) acquires a spheroidal shape as the
rotation increases. For rotation larger than a certain critical value this axisymmetric configuration is
unstable and a two-lobed configuration forms. This suggests that highly rotating black holes are unstable. In
four dimensions, the Kerr bound is small enough to avoid the development of such instabilities. However, in
dimensions higher than six, rotating Myers-Perry black holes \cite{myersperry} have no Kerr-like bound, and the
instability might well set in. Recent arguments by Emparan and Myers \cite{myersemparan}, using gravitational
physics arguments, lend further support to this claim.
For rotations larger than the above mentioned critical value, the axisymmetric configuration is unstable, but
it still is an equilibrium solution. It was further shown in \cite{Cardoso:2006sj} that for even higher
rotation rates the axisymmetric family eventually goes over to a torus or ring-like configuration. This has the
natural gravitational counterpart in the black ring solution \cite{emparanreall}. An obvious question is
whether or not the black ring solution is stable. Computations in the dual fluid model are straightforward:
these solutions are unstable.

\section*{Where do we go from here?}

There is solid evidence that the correspondence surface gravity $\rightleftarrows$ surface tension allows one
to study a plethora of phenomena related to gravitational physics in higher dimensions. The wish list for a
successful completion of this programme includes: (i) a proof of the correspondence. One such proof exists for
black holes in four dimensions, but the analog membrane is rather unusual, with negative bulk viscosity and a
non-vanishing shear viscosity. The examples worked out so far hint at something slightly more surprising: as
one climbs up in spacetime dimension the analogy we described in this essay seems to work better and better,
and thus it seems like surface tension is the {\it only} ingredient that needs to be considered. (ii)
Confirmation of the predictions of the dual fluid model. Some of these, namely rotation effects and the
existence of multi-black holes, have already been confirmed, but progress in the gravitational sector is slow.
The ultimate goal is of course to use the dual model as a guiding tool, to go hand in hand with its
gravitational cousin. That is precisely the usefulness of analogies.

\medskip
\centerline{\bf Acknowledgments} \medskip
We would like to thank the many suggestions by the participants of the KITP Program ``Scanning New Horizons: GR
Beyond 4 Dimensions", of the ``$16$th Midwest Relativity Meeting'', and of the Hebrew University programme:
``Einstein's Gravity in Higher Dimensions". In particular, we warmly thank Emanuele Berti, Luca Bombelli, Marco
Cavagli\`a, Roberto Emparan, Barak Kol, Troels Harmark, Jordan Hovdebo, Akihiro Ishibashi, Donald Marolf,
Robert Myers, Niels Obers, Leonard Parker, Michael Seifert and Toby Wiseman for valuable comments and
suggestions, that eventually lead to a solid picture of how this correspondence works. This work was partially
funded by Funda\c c\~ao para a Ci\^encia e Tecnologia (FCT) –- Portugal through projects PTDC/FIS/64175/2006
and POCI/FP/81915/2007. OD acknowledges financial support provided by the European Community through the
Intra-European Marie Curie contract MEIF-CT-2006-038924.

\bibliography{paper}

\end{document}